# Deep Learning Based Dominant Index Lesion Segmentation for MR-guided Radiation Therapy of Prostate Cancer

**Josiah Simeth, Jue Jiang, Anton Nosov, Andreas Wibmer, Michael Zelefsky, Neelam Tyagi\*, Harini Veeraraghavan\***

*(equal senior authors)

## ABSTRACT

### Background

Dose escalation radiotherapy enables increased control of prostate cancer (PCa) but requires segmentation of dominant index lesions (DIL). This motivates the development of automated methods for fast, accurate, and consistent segmentation of PCa DIL.

### Purpose

To construct and validate a model for deep-learning-based automatic segmentation of PCa DIL defined by Gleason score (GS) $\geq$3+4 from MR images applied to MR-guided radiation therapy. Validate generalizability of constructed models across scanner and acquisition differences.

### Methods

Five deep-learning networks were evaluated on apparent diffusion coefficient (ADC) MRI from 500 lesions in 365 patients arising from internal training Dataset 1 (156 lesions in 125 patients, 1.5Tesla GE MR with endorectal coil), testing using Dataset 1 (35 lesions in 26 patients), external ProstateX Dataset 2 (299 lesions in 204 patients, 3Tesla Siemens MR), and internal inter-rater Dataset 3 (10 lesions in 10 patients, 3Tesla Philips MR). The five networks include: multiple resolution residually connected network (MRRN) and MRRN regularized in training with deep supervision implemented into the last convolutional block (MRRN-DS, Unet, Unet++, ResUnet, and fast panoptic segmentation (FPSnet) as well as fast panoptic segmentation with smoothed labels (FPSnet-SL). Models were evaluated by volumetric DIL segmentation accuracy using Dice similarity coefficient (DSC) and the balanced F1 measure of detection accuracy, as a function of lesion aggressiveness and size (Dataset 1 and 2), and accuracy with respect to two-raters (on Dataset 3). Upon acceptance for publication segmentation models will be made available in an open-source GitHub repository.

### Results

In general, MRRN-DS more accurately segmented tumors than other methods on the testing datasets. MRRN-DS significantly outperformed ResUnet in Dataset2 (DSC of 0.54 vs. 0.44, p<0.001) and the Unet++ in Dataset3 (DSC of 0.45 vs. p=0.04). FPSnet-SL was similarly accurate as MRRN-DS in Dataset2 (p = 0.30), but MRRN-DS significantly outperformed FPSnet and FPSnet-SL in both Dataset1 (0.60 vs 0.51 [p=0.01] and 0.54 [p=0.049] respectively) and Dataset3 (0.45 vs 0.06 [p=0.002] and 0.24 [p=0.004] respectively). Finally, MRRN-DS produced slightly higher agreement with experienced radiologist than two radiologists in Dataset 3 (DSC of 0.45 vs. 0.41).

### Conclusions

MRRN-DS was generalizable to different MR testing datasets acquired using different scanners. It produced slightly higher agreement with an experienced radiologist than that between two radiologists. Finally, MRRN-DS more accurately segmented aggressive lesions, which are generally candidates for radiative dose ablation.

## INTRODUCTION

In recent years MRI has gained prominence in prostate cancer (PCa) radiotherapy. MRI soft tissue characterization and multimodal acquisitions allow increased imaging resolution[1], thus enabling more sophisticated radiotherapy techniques such as external beam radiotherapy for precise targeting of the tumors within the prostate gland. Nevertheless, local failures often occur at the site of the original index lesion[2,3]. The dominant index lesion (DIL) is defined as the largest lesion harboring the most biologically aggressive (determined using Gleason score) cancer and occurring in more than 80% of cases with multifocal disease[4]. Boosting dose to the DIL reduces local biochemical recurrence and treatment side effects[5]. Importantly, the use of multi-parametric MRI (mp-MRI) to boost dose to DIL has demonstrated successful outcomes with similar toxicity profiles as conventional fractionation methods that treat the entire prostate gland[6–10]. However, accurate identification and segmentation of DIL is required for precise radiative ablation[5], needed to minimize radiotoxic doses to the healthy prostate and neighboring organs[10–12]. Currently, DIL is determined by radiologists using the prostate imaging reporting and data system (PI-RADS) criteria, which produces moderate agreement between radiologists[13]. Reliably accurate delineation of such lesions by radiologists is time consuming, difficult and impractical[8,9,14], thus creating a clinical need for fast and accurate DIL auto-segmentation methods.

Deep learning (DL)[5,15–21] and radiomics based machine learning classifiers[22–26] have shown the ability to classify PCa aggressiveness characterized by Gleason scores (GS). Many works using MRI have predominantly focused on segmenting the entire prostate gland rather than focal dominant lesions[5,15,16,27–35], because the whole prostate gland is treated to the same prescription. Few recent works studied the problem of intra-prostatic lesion segmentation from MRI[5,18,20,36,37], though whole prostate gland segmentation was often combined as an additional task to guide lesion detection and segmentation. Importantly, the afore-mentioned methods typically used larger and more aggressive lesions with Gleason scores exceeding 6 or PIRADS score exceeding 4[17,20,37]. Relatedly, except for Duran et al[37] and Dai et al[5], single institution or multi-institution datasets with similar scanner acquisitions were used for evaluating the AI models, which does not indicate the model generalizability with different MR acquisitions. Our contribution improves upon prior works by (i) evaluating accuracy of PCa segmentation with different lesion aggressiveness, lesion sizes, and locations (peripheral vs. transition zone) using multiple DL methods, (ii) evaluating the generalization accuracy using testing sets with entirely different MRI scanners and protocols than the training dataset (training on 1.5T GE MR, and testing on 1.5T GE MR, 3T Siemens MR, and 3T Philips MR simulator). A total of 365 MRI scans were analyzed. (iii) We also conducted a preliminary evaluation of the DL segmentations with respect to two radiologists to measure consistency. (iv) Finally, we optimized and improved a multiple resolution residual network (MRRN) specifically developed for segmenting lung tumors of varying sizes in CT[38], with deep supervision (MRRN-DS) to regularize network training applied to DIL segmentations from apparent diffusion coefficient (ADC) MRI. We compared MRRN-DS against representative DL models to assess accuracy improvement and performed a meta-comparison against representative published DIL segmentation methods.

## METHODS

### Datasets

Three different datasets were used in the analysis and are summarized in Table 1.

Training and testing Dataset 1 (Internal): MRI studies acquired prior to prostatectomy from 151 men diagnosed with biopsy-proven prostate cancers who later underwent prostatectomy as part of an IRB approved study. This dataset was previously used for classifying aggressive prostate cancers[39] using GS, a measure of lesion aggressiveness based on the two most common cell patterns in the lesion[40,41]. Two-dimensional axial EPI-based diffusion weighted imaging (DWI) (TR/TE = 2,500–7,700/83.3–143.5 ms, FOV = 14–24 cm, voxel size = 0.625x0.625x3 mm$^3$ to 0.78x0.78x3.6 mm$^3$, b values of 0 and 1000 s/mm$^2$) were acquired from a 1.5 Tesla GE scanner using an endorectal coil. Apparent diffusion coefficient (ADC) maps were generated by performing a pixel-by-pixel mono-exponential fit to b-values and used as input to the DL models. ADC was used to make use of the significant reduction in diffusivity characteristic of PCa compared to normal prostate [42]. The dataset was randomly split into cross-validation training (156 lesions in 125 patients, 130 lesions with GS $\geq$ 3+4) and held out testing (35 lesions in 26 patients, 27 lesions with GS $\geq$ 3+4).

Lesion contours were delineated by an expert radiologist (AW), using T2 weighted images and the ADC maps in consensus with two pathology research fellows using whole-mount step section pathological tumor maps from surgically excised prostate glands as described in a previous study[39]. Up to 5 lesions were delineated per patient. Lesion specific GS were available from the pathology maps. All but 5 lesions were categorized as located in either the peripheral zone (PZ) or transition zone (TZ).

Testing Dataset 2 (External): The ProstateX grand challenge dataset[43,44], a retrospective set of prostate MR studies including ADC maps was used in the analysis. MR scans were acquired on two Siemens 3 Tesla MR scanners, namely Magneton Trio and Skyra. Volumetric lesion segmentations made available publicly by Cuocolo et al[45] were used after performing a lesion-by-lesion quality check (299 lesions in 204 patients, 76 lesions with GS ≥ 3+4). For this dataset the lesions were categorized as located in the PZ, TZ, or anterior stroma (AS).

Testing Dataset 3 (Internal): Ten MRI scans prior to radiotherapy treatment were used. Scans included multi b-value (10 b-values between 0 and 1000 s/mm2) and EPI-based DWI (TR/TE =3700-5400/70-70.77 ms, FOV = 18 cm, voxel size = 1.125x1.125x4 mm$^3$) of the prostate acquired on a 3 Tesla Philips MR simulator using phased-array coils. Lesion designations were obtained by an expert radiologist with 15 years experience (R1) and a second radiologist (R2) with 5 years experience. All but one lesion was categorized as located in either the TZ or PZ.

**Table 1. Description of the patient and PCa characteristics for the datasets used in this study.**

| Variable | Dataset 1: Internal | Dataset 2: External ProstateX | Dataset 3: Internal |
|---|---|---|---|
| **Purpose** | Model building via cross-validation, and hold-out testing | Testing | Testing and inter-rater consistency evaluation |
| **Scanner details** | 1.5T GE MR | 3T Siemens MR | 3T Philips MR sim |
| **Total Patients** | 151 | 204 | 10 |
| **Segmented Lesions** | 191 | 299 | 10 |
| **Gleason Scores** | | | |
| **6** | 34 | 36 | 0 |
| **3+4** | 112 | 41 | 3 |
| **4+3** | 26 | 20 | 7 |
| **≥8** | 19 | 15 | 0 |
| **Unlabeled, sub-clinical** | 0 | 187 | 0 |
| **Zone** | | | |
| **Peripheral Zone** | 143 | 50 | 7 |
| **Transition Zone** | 43 | 17 | 2 |
| **Anterior Stroma** | N/A | 45 | N/A |
| **Other/unlabeled** | 5 | 187 | 1 |
| **Lesion Sizes (cm$^3$)** | | | |
| **Median [IQR]** | 0.8 [0.4 - 1.5] | 1.1 [0.5 - 2.4] | 0.6 [0.5 - 1.1] |

## Data preparation

Images from all three datasets were resampled at 0.625x0.625x3 mm$^3$ and cropped to 128x128 image regions enclosing the prostate. Any mislabeled pixels defined as isolated single-pixel lesions in Dataset 2 were removed automatically through image processing used to detect small and connected cancer regions (number of voxels < 2).

**Deep learning networks architecture**

Five different architectures were investigated: (1) a Unet (2) a multi-resolution residual network (MRRN), and an extension of the MRRN with deep supervision (MRRN-DS), (3) Unet++[46], (4) Deep residual Unet (ResUnet[47]), and (5) a modified Fast Panoptic Segmentation network (FPSnet), both without and with smoothed labels (FPSnet and FPSnet-SL). Architecture diagrams are shown in Figure 1.

**Figure 1.** The architecture for (a) Unet, (b) Unet++, (c) ResUnet, (d) MRRN/MRRN-DS, and (e) FPSnet models. (MRRN matches MRRN-DS except for the removal of the deep supervision connections (red) bypassing the final RU blocks.

Unet: Unet was chosen for comparison due to the frequent use of this method by several prior works. Our implementation (see Figure 1a) followed prior work [48], wherein 4 max-pooling layers were used to downsample features in the encoder to model different image resolutions and 4 up-pooling layers to resample the features in the decoder and back into the full image resolution level. Skip connections were used to concatenate features from the encoder with the corresponding feature layers (image resolution level) in the decoder layers. Convolution blocks were followed by batch normalization and ReLU layers in each feature layer. This model had a total of 13 million parameters.

Unet++: The Unet++[46] improves upon the Unet with the addition of a dense "nested Unet" architecture with deep supervision. Deep supervision is used to enhance backpropagation during training and provide better regularization of network training, especially when using deeper networks[49]. The nested structure is such that the downpooling encoding and upsampling decoder structure is nested throughout the model (see Figure 1b). In this case the nesting is such that the first convolution block is effectively the first encoder for a 2-layer Unet with the next encoder, a 3-layer Unet with the next 2 encoders, a 4-layer Unet with the next 3 layers, and finally a 5-layer Unet with the entire encoder line. The outputs of each of these nested Unets is then used for deep supervision. This implementation had a total of 9 million parameters.

Res-Unet: The Res-Unet[47] improves upon the Unet through the replacement of the simple convolution blocks with residual units (compare the Unet to the ResUnet in Figure 1). These residual units include skip connections to aid backpropagation in very deep networks. This implementation had a total of 32 million parameters.

MRRN and MRRN-DS: MRRN[38] combines aspects of both densely connected and residual network architectures by simultaneously combining features extracted at different image resolutions (obtained through pooling operations across the feature layers) and residual connections between adjacent feature levels. This approach has been shown to provide more accurate segmentation of CT scans for lung tumors[38] as well as normal organs than Unet architecture[50]. Similar to the Unet, a contraction path composed of feature encoder layers and produced through feature max pooling (4 layers are used) are combined with decoder layers (4 unpooling layers) to extract the segmentation features and generate the final segmentation. However, instead of only concatenating features from encoder with the decoder from the matching resolution, MRRN uses residual connections from prior feature layers as well as feature concatenation with features computed at multiple image resolutions simultaneously. The updated features are passed back into the residual feature streams used to provide refined information for subsequent feature layers. Thus, this network simultaneously combines residual connections and dense connections to improve the network's capacity. Again, note that unlike the residual connections used in ResUnet, where the residual connections are only used to concatenate features from the prior layer with the current layer inputs, the MRRN modifies the residual feature streams themselves with the outputs of individual layers such that the feature activations are modulated with increasing network depth. Further, while densely connected networks such as Unet++ feed forward activations from multiple prior layers as a concatenation of features they do not explicitly modulate the features carried by multiple feature streams to contextualize the activations to be used in deeper layers as is done in the MRRN.

MRRN-DS improves upon MRRN using deep supervision to enhance backpropagation during training and provide better regularization of network training[49]. Deep supervision was provided by using the output of the penultimate convolutional block before the connection to the residual unit combining the first feature stream from the input (see Figure 1d). This supervision depth was selected experimentally by varying the layer at which supervision was performed and selecting the depth that maximized DSC accuracy in the validation set (see supplementary Table S1). The loss function is thus implemented as:

$$L_{combined}(\hat{y}, y) = \mu L(\hat{y}, y) + (1 - \mu)L(\hat{y}_{deep\_sup}, y), \qquad [1]$$

where $L(\hat{y}, y)$ is the loss function, y is the ground truth, $\hat{y}$ and $\hat{y}_{deep\_sup}$ are predictions from the network output and from the internal layer provided with deep supervision respectively, and $\mu$ is a weighting term. Both MRRN and MRRN-DS had 39 million parameters. The weighting term was selected as 0.75 experimentally. The value was determined by varying $\mu$ from 0.5 to 0.95 and selecting the optimal DSC cross-validation result (see supplementary Table S2).

FPSnet: We enhanced the FPSnet from Geus et al [51] to combine PCa detection using bounding box regression followed by dense pixel prediction to extract PCa segmentation. The original network, developed for classification combined bounding box regression with instance classification. This network employed feature extraction performed by a pretrained backbone (in this case ResNet-50 trained on the ImageNet database [52] ) to generate convolutional feature maps, which are then used to feed sub-networks performing PCa detection using RetinaNet[53] and segmentation. Our implementation replaced the segmentation head used in FPS with a Unet architecture to allow multi-scale feature extraction and integration (see Figure 1e). Two variations of the FPSnet model were implemented. One model called FPSnet-SL for FPSnet with smoothed labels (resulting in real-valued labels ranging between 0 to 1) was trained with smoothed labels produced through bi-linear interpolation of the expert segmentation masks to resample images to 256 x 256 pixels. The second model called FPSnet was trained with binary labels by resampling the segmentation masks using nearest neighbor interpolation to 256 x 256 pixels.

**Implementation**

All models were implemented using Pytorch[54]. All networks were trained using Adam[55] and a learning rate of 1e-4 for the first 20 epochs and a linearly decaying learning rate for an additional 100 epochs. The batch size was fixed to 3. Online data augmentation was performed through left-right flips, scaling, rotation, and elastic deformation. All segmentation models were trained via a five-fold cross validation with identical training datasets and optimized using a soft Dice loss. Early stopping was used to prevent overfitting. For each architecture, the model with best validation results was used for testing. FPSnet was trained with 3 channels of ADC including both the relevant slice and adjacent slices for 3-channel input to the pretrained ResNet-50. All other models used 5 channels of ADC including two adjacent slices on either side of the relevant slice (superior and inferior). The input to FPSnet-SL was bilinearly upsampled from 128x128 to 256x256 for input to the pretrained ResNet-50, which created smoothed training labels resembling the training process for label softening methods[56,57]. The training process was repeated with binary labels produced using nearest neighbor interpolation for standard training (FPSnet).

**Evaluation metrics**

Model performances were measured in terms of recall, precision, F1, and DSC of lesions with respect to ground truth (or expert segmented) DIL segmentations.

$$Recall = \frac{TP}{P} \qquad\qquad [2]$$

$$Precision = \frac{TP}{TP + FP} \qquad\qquad [3]$$

$$F1 = \frac{2 \times Precision \times Recall}{Precision + Recall} \qquad\qquad [4]$$

TP is the number of true positives (correctly detected lesions), FP is false positives (incorrectly identified lesion), and P is the number of actual lesions.

DSC was computed as

$$DSC = \frac{2TP}{2TP + FP + FN} \qquad\qquad [5]$$

for all lesions overlapping with the expert segmented lesions, where FN is false negatives. Only lesion volumes with a DSC overlap exceeding 0.1 with the expert delineation were considered as true detections as previously done by Saha et al [17] and Mckinney et al [58]. Segmented lesions with volumes under 0.1 cm$^3$ were ignored as negligible.

**Statistical analysis**

Statistical comparison between the various methods was performed using paired, two-sided Wilcoxon signed-rank tests with the DSC metric for the testing sets. Corresponding effect sizes were calculated using the matched biserial coefficient[59]. Analysis of differences in the segmentation accuracies measured using DSC for the individual methods with respect to the GS, lesion sizes, and PCa location in the prostate gland was measured using unpaired and two-sided Wilcoxon rank sum test. All statistical analysis was performed using Matlab version R2021b and only p-values with p < 0.05 were considered statistically significant.

**Experiments:**

**Segmentation accuracy dependence to lesion characteristics:** The dependence of segmentation accuracy on the lesion aggressiveness for low risk lesions (GS = 3+3) vs. intermediate risk lesions (GS = 3+4) vs. high risk lesions (GS ≥ 4+3) was evaluated for both Dataset 1 and 2. Segmentation accuracy with respect to tumor location (peripheral zone vs. transition zone or anterior stroma) was assessed for both Dataset 1 and 2. Finally, accuracy with respect to lesion size (< 1 cm$^3$ vs. 1 to 2 cm$^3$ vs. > 2 cm$^3$) was computed for both Dataset 1 and 2.

**Segmentation accuracy with respect to two raters:** The DL segmentations were compared against the more experienced of the two radiologists (AW) in order to determine if the segmentation variability was within the variability of two raters[5,60,61] as a measure of the utility of the DL methods.

**MRRN structural ablation:** To better understand the contributions of the feature streams of the MRRN we evaluated network performance through structural ablation. MRRN segmentation accuracy was assessed with removal of the full resolution features stream (see residual stream 1 in supplementary Figure S1, corresponding to layer 1), and removal of all feature streams except for the full resolution feature stream (feature streams 2, 3 and 4 in supplementary Figure S1, corresponding to layers 2, 3, and 4).

The implementation of deep supervision was also assessed by varying the decoding layers receiving supervised losses. This was accomplished by computing the deep supervision loss (Equation 1) using the upsampled outputs generated from the decoder layers (see supplementary Figure S1 for labeled supervision structure).

 Hyperparameter optimization was also performed to determine the choice of weighting term **μ** (ranging from 0.5 to 0.95) used to weight the deep supervision losses in the MRRN-DS network.

**RESULTS**

**Accuracy comparisons of multiple network architectures on internal and external datasets with different MRI acquisition parameters:**

Segmentation and lesion detection accuracies for aggressive lesions (GS ≥ 7) are shown in Table 2 for the Dataset 1 using held out testing data, and testing with the external Dataset 2 and Dataset 3. Cross tabulated Wilcoxon rank sum tests measuring differences in the accuracies as p-values and effect sizes reported as matched biserial coefficient are provided in supplementary Table S3.

As shown in Table 2, MRRN and MRRN-DS produced the most accurate tumor segmentations (median DSC of 0.60 each) for the withheld testing 1.5T GE data (Dataset 1). All analyzed networks including Unet, Unet++, MRRN, and MRRN-DS significantly outperformed FPSnet, with MRRN and MRRN-DS having the greatest effect sizes of 0.55 and 0.54 respectively (supplementary Table S3). MRRN-DS was also significantly more accurate than FPSnet-SL (p = 0.049). FSPnet-SL and FPSnet were similarly accurate (p = 0.24).

In the Dataset 2, although FPSnet-SL produced the best median DSC of 0.61, this accuracy was not significantly higher than MRRN-DS ($p = 0.055$), indicating a minor improvement. ResUnet was the least accurate method for this dataset with all methods significantly outperforming this method (supplementary Table S3). Contrary to Dataset 1, FPSnet-SL significantly outperformed FPSnet ($p < 0.001$).

In the Dataset 3 scanned with 3.T Philips scanner, MRRN-DS (median DSC of 0.45) was more accurate than FPSnet-SL and it was similarly accurate as the ResUnet (Table 2). The Unet, ResUnet, MRRN, MRRN-DS, and FPSnet-SL were all more accurate than the FPSnet. Similar to the results in Dataset1, MRRN-DS was significantly more accurate than both FPSnet ($p = 0.002$) and FPSnet-SL methods ($p = 0.0039$) and had higher biserial coefficient with respect to FPSnet of 1.0 compared with Unet and ResUnet methods and FPSnet.

Detection accuracy measured using F1 scores are also reported for completeness in Table 2 as are the detection and segmentation accuracies for all lesions including those with GS = 6 are included in the supplementary Table S4, along with cross-validation results. For both Dataset 1 and Dataset 2 all models saw slight drops in recall or median DSC when incorporating lesions with GS = 6. Detection statistics for lesions from Dataset 2 without Gleason scores are given in supplementary Table S7.

**Table 2.** Segmentation DSC, recall, precision, and F1score for DIL lesions (GS ≥ 7) for all datasets. Median DSC and ranges are calculated across all DIL lesions. Column headings indicate the total number of DIL lesions (GS ≥ 7) in the given patient dataset, along with the median lesion size for this sub population. DSC values that were significantly superior (p<0.05) to at least one other model are indicated by an Asterix (*), while models significantly worse are indicated by a negated Asterix (*). Note that MRRN-DS is the MRRN with deep supervision, and FPSnet-SL is FPSnet trained using smoothed lesion labels.

| | Dataset 1 | Dataset 2 | Dataset 3 |
|---|---|---|---|
| | (held-out testing) | (external testing) | (additional testing) |
| | (n lesions = 27)  (median 1.1 cc) | (n=76)  (median 1.1 cc) | (n=10)  (median 0.6 cc) |
| **Model** | Lesion DSC, Median (IQR) | Lesion DSC, Median (IQR) | Lesion DSC, Median (IQR) |
| **Unet** | 0.54 (0.45-0.70)* | 0.50 (0.17-0.65)* | 0.31 (0.00-0.53)* |
| **Unet++** | 0.58 (0.43-0.69)* | 0.54 (0.12-0.69)* | 0.23 (0.00-0.48)-* |
| **ResUnet** | 0.54 (0.35-0.69) | 0.44 (0.00-0.65)-* | **0.45 (0.24-0.52)**\* |
| **MRRN** | **0.60 (0.48-0.69)**\* | 0.49 (0.27-0.65)* | 0.38 (0.00-0.54)* |
| **MRRN-DS** | **0.60 (0.49-0.71)**\* | 0.54 (0.23-0.68)* | **0.45 (0.22-0.54)**\* |
| **FPSnet** | 0.51 (0.21-0.67)-* | 0.48 (0.14-0.63)-* | 0.06 (0.00-0.16) -* |
| **FPSnet-SL** | 0.54 (0.31-0.66)-* | **0.61 (0.30-0.72)**\* | 0.24 (0.00-0.29)-*/* |
| | **Precision, Recall, F1** | **Precision, Recall, F1** | **Precision, Recall, F1** |
| **Unet** | 0.49, 0.96, 0.65 | 0.30, 0.79, 0.43 | 0.32, 0.70, 0.44 |
| **Unet++** | 0.48, 0.96, 0.64 | 0.27, 0.78, 0.40 | 0.26, 0.70, 0.38 |
| **ResUnet** | 0.47, 0.93, 0.63 | 0.23, 0.71, 0.34 | 0.32, 0.90, **0.47** |
| **MRRN** | 0.45, 0.96, 0.61 | 0.27, 0.82, 0.41 | 0.25, 0.70, 0.37 |
| **MRRN-DS** | 0.49, 1.00, 0.66 | 0.26, 0.80, 0.39 | 0.23, 1.00, 0.37 |
| **FPSnet** | 0.61, 0.81, **0.69** | 0.29, 0.78, 0.42 | 0.21, 0.50, 0.30 |
| **FPSnet-SL** | 0.42, 0.89, 0.57 | 0.29, 0.84, **0.44** | 0.27, 0.70, 0.39 |

Figure 2 shows representative segmentations produced by the various methods from the testing datasets. As shown, MRRN-DS closely followed expert delineations with fewer over-segmentations compared to other models. Segmentation was more accurate for lesions with lower ADC, as evidenced by a moderate Spearman correlation of $\rho$ = -0.51 (p<1e-13) between the MRRN-DS lesion DSC and the ground truth median ADC in each lesion in Dataset 1. A weaker, but still significant correlation was seen for Dataset 2 $\rho$ = -0.26 (p=0.006). DSC also tended to be greater for single lesions as compared to multifocal lesions.

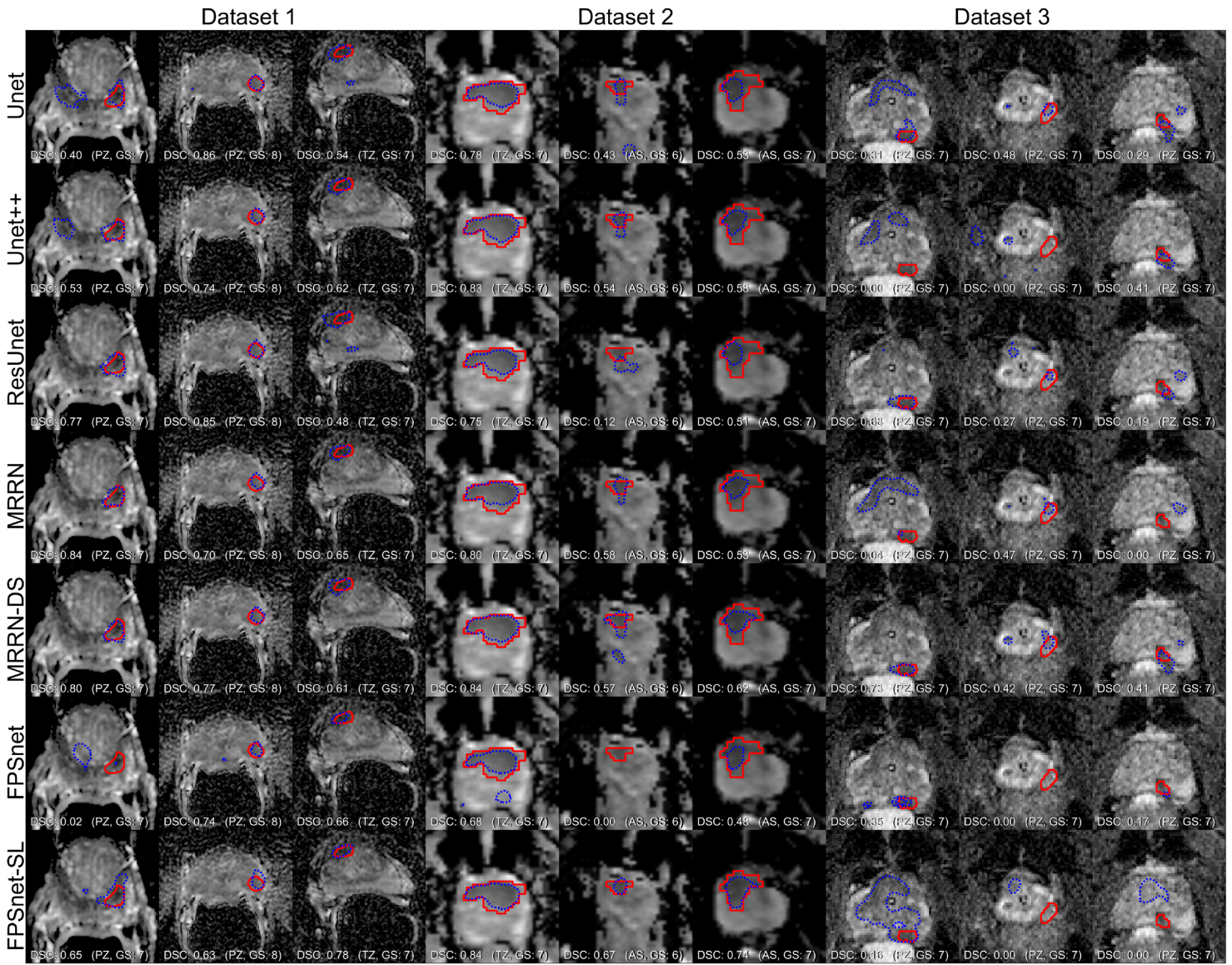

**Figure 2. Representative segmentations produced by the various models in each test set (first three columns from Dataset 1, next three columns from Dataset 2, last three columns from Dataset 3). Expert segmentations are delineated by the solid red line. Dashed blue contours correspond to algorithm segmentations. The slice DSC, lesion zone, and lesion GS is given for each image.**

**Segmentation accuracy differences in relation to lesion aggressiveness, location, and size:**

Figure 3, Figure 4, and Figure 5 show the segmentation accuracy measured using the DSC metric produced by the various models for various GS groups (3+3 versus 3+4 versus 4+3 or greater, Figure 3), lesion sizes (small [<1cc] versus medium [1cc to 2cc] versus large [> 2cc], Figure 4), and lesion locations (PZ, TZ, and AS, Figure 5). Cross-validation results were combined with held-out testing sets in Dataset

1 in order to obtain a larger number of cases for visualizing the results for the disaggregated analysis. Disaggregated results for the withheld testing alone can be found in supplementary Figure S2.

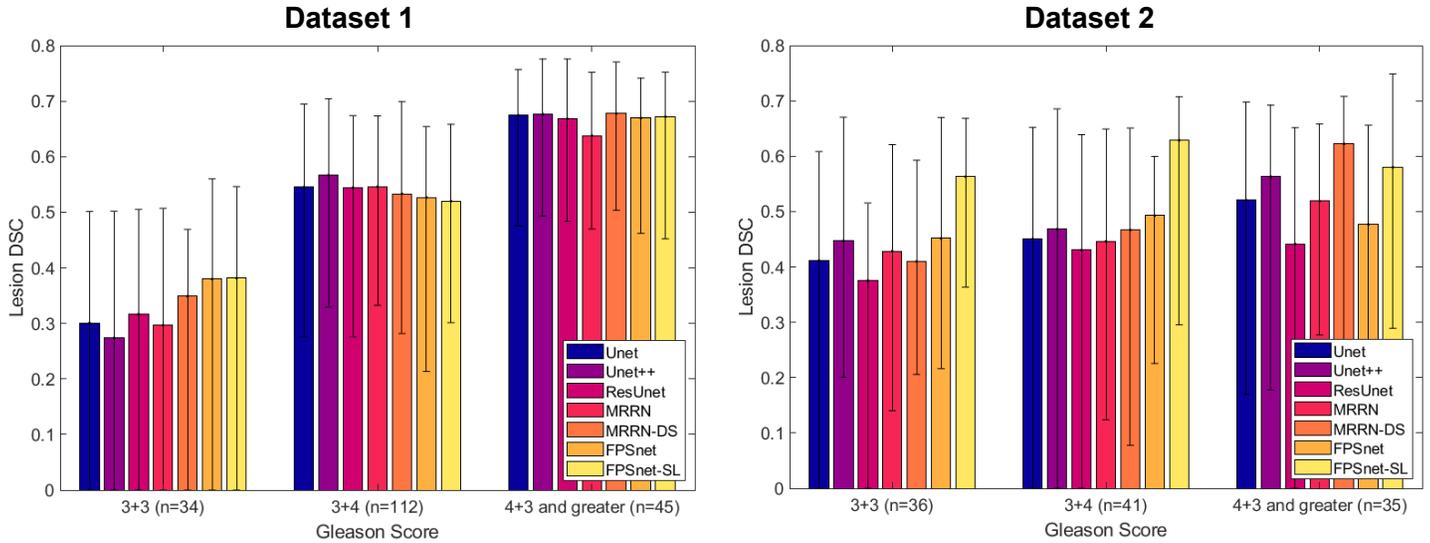

**Figure 3.** Comparison of segmentation results by the analyzed methods for low risk (GS of 3+3), intermediate risk (GS of 3+4) and high risk (GS of 4+3 and higher) lesions. Each bar represents the median lesion DSC, with error bars indicating the interquartile range. The left plot shows the results of the Dataset 1 (Cross-validation + held-out testing), while the right plot shows Dataset 2 (ProstateX).

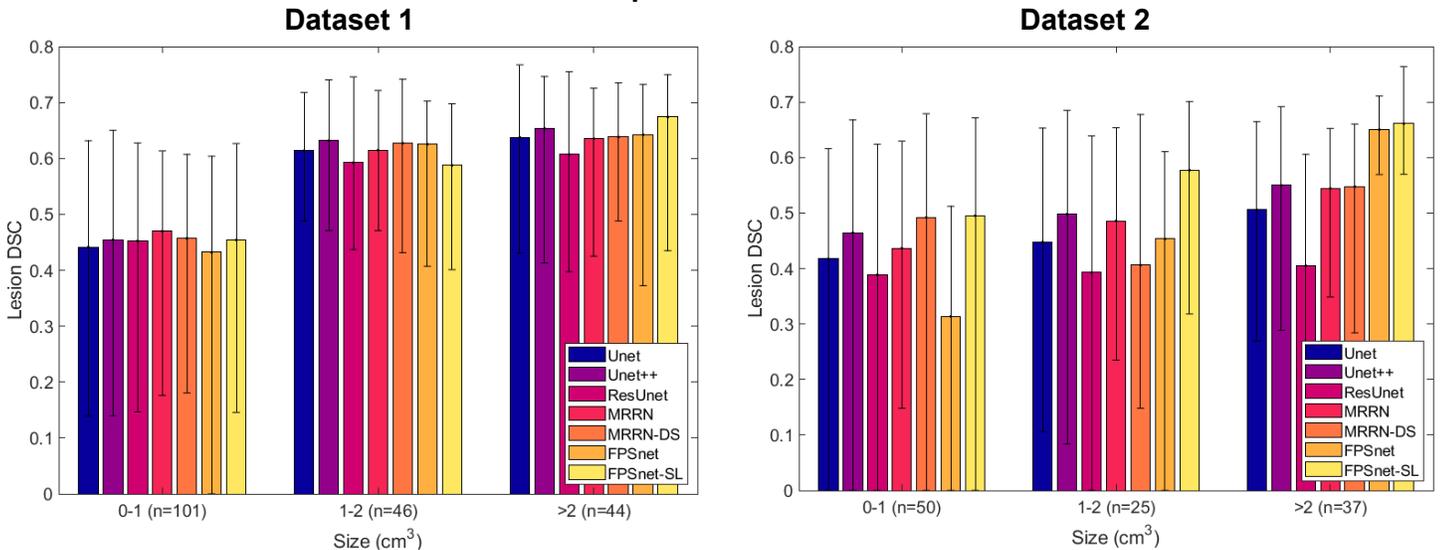

**Figure 4.** Comparison of segmentation results by the analyzed methods for differing volume. Each bar represents the median lesion DSC, with error bars indicating the interquartile range. The left plot shows the results of the Dataset 1 (Cross-validation + held-out testing), while the right plot shows Dataset 2 (ProstateX).

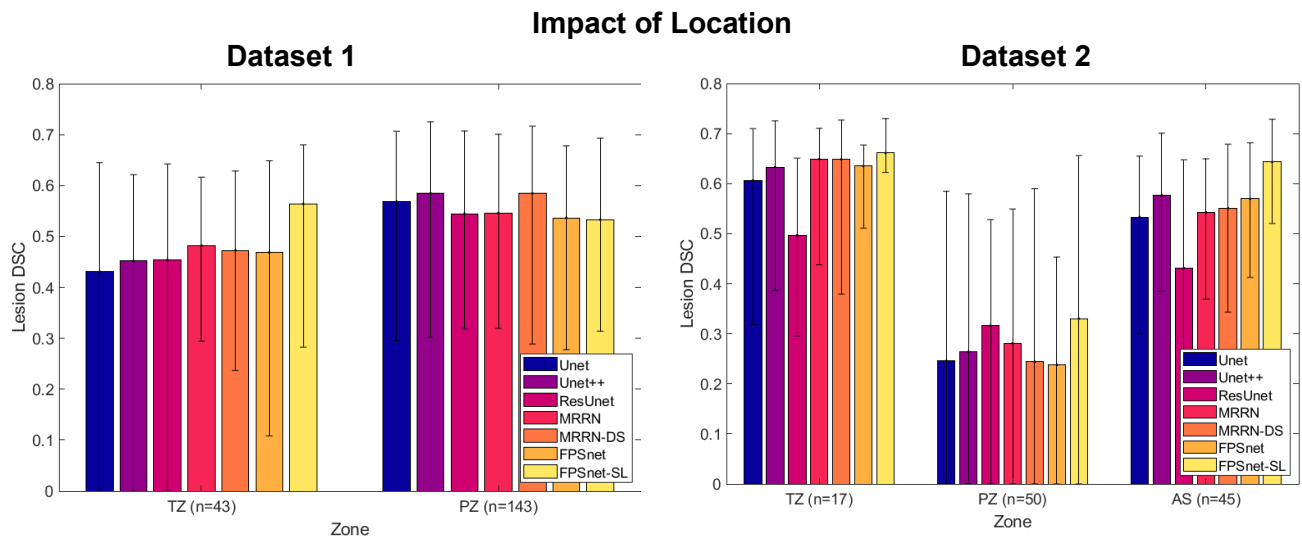

**Figure 5. Comparison of segmentation results by the analyzed methods for differing tumor locations. Each bar represents the median lesion DSC, with error bars indicating the interquartile range. The left plot shows the results of the Dataset 1 (Cross-validation + held-out testing), while the right plot shows Dataset 2 (ProstateX). TZ = transition zone; PZ=peripheral zone; AS= anterior stroma.**

**Model comparisons based on lesion characteristics**

When comparing the models to one another across the various groups of lesions in terms of size, aggressiveness, and zone, no single model was superior across all lesion characteristics and datasets.

MRRN-DS produced the most accurate segmentations for the most aggressive lesions in both Dataset 1 and Dataset 2 (GS ≥ 4+3). FPSnet and FPSnet-SL were among the least accurate models in Dataset1 and Dataset3 for aggressive lesions. However, both models provided more accurate segmentations compared to all other methods for the less aggressive GS = 6 lesions in both Dataset 1 and Dataset 2. FPSnet-SL also achieved higher DSC to all other models for lesions of at least 2cc in the Dataset 2, albeit this method was overall similarly accurate as the MRRN-DS method (p = 0.055).

**Segmentation accuracy by Gleason score**

In Dataset 1, all models performed significantly better for high risk compared to intermediate risk lesions (GS ≥ 4+3 vs 3+4). Median DSC values for lesions grouped by Gleason score with Wilcoxon median test p-values are recorded in supplementary Tables S5 and S6.  On the other hand, when comparing the low and intermediate risk lesions (GS of 3+4 and GS of 3+3), only FPSnet and FPSnet-SL showed similar accuracy for these lesion types. All other models produced significantly less accurate segmentation for GS of 3+3 lesions. (see supplementary Table S5). However, in Dataset 2, there was no difference in accuracy for the various models between the highly aggressive vs. Intermediate (GS of 3+4) vs. less aggressive (GS of 3+3) lesions (see supplementary Table S6).

**Segmentation accuracy by lesion volume**

In Dataset 1, all models segmented medium sized lesions (1cc to 2cc) with higher accuracy than small lesions (< 1cc). Model median DSC values for lesions grouped by volume with Wilcoxon median test p-values are recorded in supplementary Tables S5 and S6. No model showed a significant difference in segmentation accuracy between medium (1cc to 2cc) and large (>2cc) lesions. In Dataset 2, only FPSnet and FPSnet-SL, were significantly more accurate for medium lesions compared to small lesions (supplementary Table S6), and no model showed a significant accuracy difference between medium and large lesions.

**Segmentation accuracy by lesion location**

In Dataset 1, MRRN-DS, Unet, Unet++, and ResUnet produced significantly more accurate segmentations of tumors occurring in the peripheral zone (PZ) than those occurring in the transition zone (TZ). Model median DSC values for lesions grouped by location with Wilcoxon median test p-values are recorded in supplementary Tables S5 and S6. FPSnet and FPSnet-SL were similarly accurate for lesions occurring in PZ and TZ (see Supplementary Table S5). The lesions in the peripheral zone had a median volume of 0.74 cc, inter-quartile range (IQR) of 0.35 to 1.43cc, with 10% having GS 4+3 or greater, while those occurring in the transition zone had a median volume of 0.92 cc, IQR of 0.32 to 1.75cc, with 27% having a GS 4+3 or greater. Representative segmentation of lesions occurring in the PZ and TZ are shown in supplementary Figure S3.

In the Dataset 2, all models except ResUnet produced significantly higher segmentation accuracy in the TZ relative to the PZ. The lesions in the PZ had a median volume of 0.65cc, inter-quartile range (IQR) of 0.35 to 1.6 cc, with 30% having GS 4+3 or greater for this dataset. Lesions in the TZ had a median volume of 1.57 cc, IQR of 0.77 to 2.6cc, with 24% having a GS 4+3 or greater. Lesions in the anterior stroma had a median volume of 1.6cc, IQR of 0.98 to 3.2cc, with 36% having a GS 4+3 or greater. This low accuracy for PZ lesions may be due to the lack of an endorectal coil in Dataset 2.

We analyzed the datasets for false detections outside the prostate gland, because none of the methods used prostate gland segmentation to provide a location prior as used by other prior works. Only a single instance of false detection outside the prostate was observed in the held-out testing of Dataset 1 (by the Unet). All of the methods produced false detections outside prostate gland for multiple patients in Dataset 2 (Unet: 63, Unet++: 62, ResUnet: 58, MRRN: 69, MRRN-DS: 79, FPSnet: 66, FPSnet-SL: 65). Out of prostate segmentations were most common in the rectum (see supplemental Figure S4 for an example).

**Accuracy with respect to two rater segmentations:**

Segmentation and lesion detection accuracies are shown in Table 2 for the Dataset 3. Additionally, cross tabulated Wilcoxon median test significance values and effect sizes are recorded in supplementary Table S3.

Median interobserver DSC between the primary (experience interpreting MRI = 15 years) and secondary (experience = 5 years) genitourinary radiologists was a DSC of 0.41, (range: 0.33 to 0.84). The MRRN-DS and ResUnet models achieved the best DSC relative to the primary reviewer, with a median DSC of 0.45 and perfect recall for MRRN-DS. In comparison, recall for other methods were, Unet of 7/10, Unet++ of 7/10, ResUnet of 9/10, MRRN of 7/10, FPSnet of 5/10, and FPSnet-SL of 7/10. The MRRN-DS produced more accurate segmentation than the Unet++ , FPSnet, FPSnet-SL and was similarly accurate as the Unet, ResUnet, and MRRN. However, MRRN-DS model also had a high false positive rate, with an average of 1.9 false positives per lesion, while the FPSnet had the lowest false positive rate of 0.9 per lesion. The false positive rates for other models were: Unet of 1.2, Unet++ of 1.3, ResUnet of 1.1, MRRN of 1.5, MRRN-DS of 1.9, FPSnet of 0.9, and FPSnet-SL of 1.6. Figure 6 shows representative segmentations for MRRN-DS and both raters. As shown, MRRN-DS showed a close agreement with reviewer 1 for the first two cases. One case that resulted in over segmentation using this model is also shown (Figure 6, third panel). This lesion had an average ADC of 664 ($10^{-6}$ mm$^2$/s) and occurred in the TZ. In general, for the analyzed 10 cases, the MRRN-DS and ResUnet produced median DSC accuracies of 0.43 and 0.47, respectively for lesions occurring in the peripheral zone (n = 7) and 0.49 and 0.44, respectively for lesions occurring in the transition zone (n = 2). Accuracy for the other methods were lower for peripheral zone (DSC Unet: 0.21, Unet++: 0.20, MRRN: 0.38, FPSnet: 0.13, FPSnet-SL: 0.20) but similar in the transition zone for all but FPSnet (DSC Unet: 0.50, Unet++: 0.42, ResUnet: 0.44, MRRN: 0.49, FPSnet: 0.08, FPSnet-SL: 0.33).

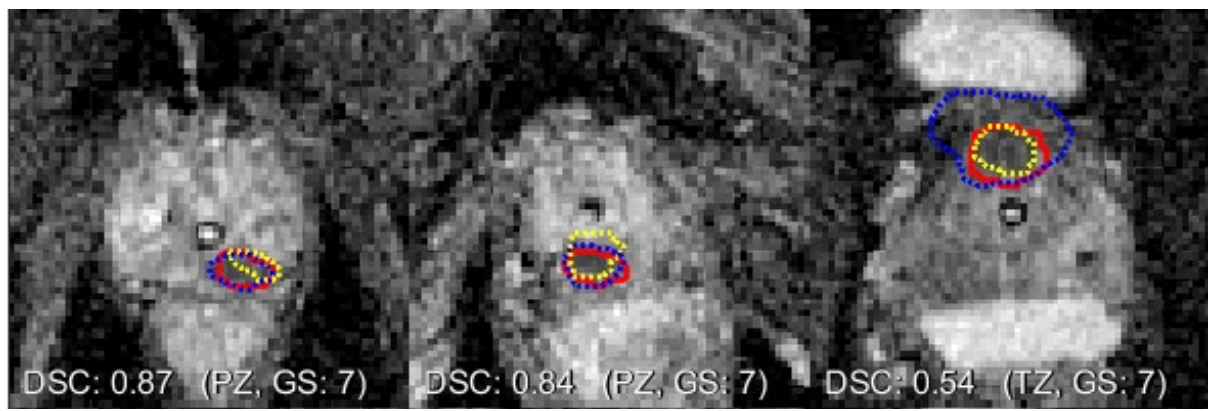

**Figure 6. Example Dataset 3 DIL delineations by the primary reviewer (red), secondary reviewer (yellow) and MRRN-DS model (blue). The first two lesions are detected and relatively well delineated by MRRN-DS, while in the final image the model over segments the lesion.**

**Ablation experiments:**

MRRN configuration had only a minor impact on segmentation performance (see table S8 in the supplement). In both Dataset1 and Dataset2 we observed slight decreases in recall rate (0.90 to 0.89 and 0.82 to 0.79) when only the full resolution feature stream corresponding to layer 1 was retained (see Figure 1 for MRRN layer designations), along with a slight increase in segmentation accuracy (DSC of 0.58 to 0.59 and DSC of 0.49 to 0.51). Similarly, removal of the layer 1 feature stream resulted in a slight increase in DSC (0.58 to 0.59 and 0.49 to 0.50), but only changed recall for Dataset2 (0.82 to 0.79).

Variation in the depth of deep supervision produced small changes in DSC, with slightly lower segmentation accuracy for supervision at the coarser resolution layers, indicating that supervision at layer 1 was most effective. DSC values can be seen in supplementary Table S1.

Variation of the hyperparameter µ gave optimal DSC for a moderate value of 0.75 with slightly lower segmentation accuracy for higher or lower weighting parameters. DSC values can be seen in supplementary Table S2.

**DISCUSSION**

A key innovation in this paper is the application of a network developed for CT images to MRI scans combined with rigorous testing of its generalization ability with a variety of MRI scanner acquisitions and protocols. The network architecture was itself improved with deep supervision and evaluations were performed analyzing the relevance of the various feature streams and supervision at various levels. Our analysis showed that the MRRN-DS trained on a cohort of patients acquired using a 1.5Tesla GE scanner with endorectal coil produced similarly accurate segmentation of DIL in testing sets consisting of ADC MRI acquired from 3.0Tesla Siemens (Magnetom and Skyra) and slightly less accurate segmentation for DILs acquired with 3.0Tesla Philips scanners. In general, MRRN-DS produced more accurate lesion segmentations than other methods on the testing datasets with no other method showing a clear accuracy improvement over all other methods in all three datasets. For instance, ResUnet was similarly accurate as the MRRN-DS in Dataset 3, whereas Unet++ was only slightly less accurate than MRRN-DS in Dataset 1. FPSnet-SL produced most accurate segmentation in Dataset 2 but was significantly worse than all other methods in Datasets 3 and significantly less accurate than MRRN-DS in Dataset 1. Overall, MRRN-DS showed an ability to provide the best accuracy in 2 out of 3 datasets with very different MR acquisitions and improved the agreement with respect to an experienced radiologist in the Dataset 3. This indicates an ability to better generalize to different MR acquisitions and validates MRRN for problems outside the Lung CT context it was developed for.

MRRN-DS also achieved as good, or better DIL segmentation accuracies compared to prior published methods (Table 3). The higher accuracy of MRRN compared to the other evaluated network architectures especially on testing datasets with different MRI acquisitions might have resulted from the higher capacity of the network because it combines features of residual and dense networks. The lack of significant performance loss in MRRN feature stream ablation was unexpected given our prior results[38] for lung cancers segmented in CT images, which showed clear improvement for the full network compared to configuration with only the full resolution feature stream. It is possible that the modest size of the training set compared to the network's capacity could have contributed to similar performance of the various network configurations.

The good performance of FPSnet in Dataset2, and very good performance of the FPSnet with smoothed training labels is interesting. The DWI series in Dataset 2 were acquired with a single-shot echo planar imaging sequence with a resolution of 2 mm in-plane and 3.6 mm slice thickness and with diffusion-encoding gradients in three directions, this results in a smoother image than our images that were acquired with much in-plane higher resolution. The additional benefit seen by smoothing the labels through bilinear interpolation (FPSnet-SL) instead of nearest neighbor interpolation (FPSnet) may indicate that further label smoothing or softening could act as a helpful regularization for generalization of PCa segmentation across differing acquisition and scanner characteristics, though additional investigation is required to determine the specific domain differences that cause the largest improvements in performance achieved by label smoothing.

To our knowledge, this is the first study performing set aside testing on MR datasets with different imaging acquisitions. Prior studies often used single institution datasets[18,36] or analyzed patient studies acquired from different institutions but using the same scanner [17,20,37]. One prior study by Dai et al [5] used datasets acquired with different scanner manufacturers but combined the two datasets to achieve accuracies similar to those of our method trained with only a single institution dataset. Table 3 shows a comparison of the various published and closely related studies to our method in terms of the analyzed datasets as well as the primary goal of the undertaken study. Different from prior studies, our approach does not require segmentation of the prostate gland to guide the detection and segmentation of the DIL lesion, thus reducing the number of steps and potential errors in the individual steps from reducing accuracy of the downstream DIL segmentation task.

Prior studies have predominantly focused on analyzing more aggressive lesions (GS > 3+3 or PIRADS > 3)[17,18,20] or larger lesions[36]. Our method achieved a DSC of 0.60 on Dataset 1 which used the same MR imaging as training set, which included both smaller and larger lesions. Eidex et al reported a DSC of 0.83 by selectively using only lesions with a DSC overlap exceeding 0.4, while Saha et al reported a comparable accuracy of 0.49 and 0.58 on internal and external testing sets using same scanners, respectively. A recent study by Gunashekar et al [62] with testing performed on MRI with whole mount prostatectomy data confirmation showed a DSC accuracy of 0.31 with an inter-rater DSC of 0.32 with the training and testing cohorts coming from internal institution albeit different magnet strengths (1.5 Tesla Siemens MRI used for training and 3 Tesla Siemens MRI used for testing). Our method produced an accuracy of 0.54 DSC on external dataset using different scanners, which is comparable to the best accuracy reported thus far on an external testing dataset by Saha et al[17]. Of note, Saha et al[17] used testing datasets provided by the same scanner manufacturer. Dai et al used two different datasets with different scanners, achieving a DSC of 0.38 when training on ProstateX data and testing on internal data [5].

We evaluated the approach to segment both less aggressive (GS = 3+3) and more aggressive lesions and found that all methods produced significantly more accurate segmentations for more aggressive lesions. Higher accuracy for larger and more aggressive lesions is desirable because they are generally candidates for radiative ablation [63] and receive greater benefit from dose escalation [64,65]. Furthermore, we analyzed accuracy with wide variation in lesion sizes, including lesions as small as 0.1 cc (median size 0.93 cc). As was expected, accuracies increased with larger lesions though not significantly for some models in Dataset 2. DSC is known to be biased positively towards larger lesions [66]. Missed detections occurred more frequently for very small lesions. For lesions >1 cc, MRRN-DS had a recall >94% (including only GS labeled lesions) though this effect was less profound in Dataset 2. Comparative analysis with other DL methods showed that MRRN and MRRN-DS were similarly accurate, which indicates that the deep supervision regularization was marginally useful. FPSnet, an approach that combines detection followed by segmentation within a detection region of interest was more accurate for less aggressive (GS = 3+3) and largest lesions on the external testing dataset. On the other hand, MRRN-DS had the best performance for the most aggressive lesions (GS≥4+3) and achieved the best performance in both Dataset 1 and Dataset 3. Dataset 3 was sourced from patients treated with hypofractionated MR-guided RT at our institution, and MRRN-DS achieved a closer agreement than the two radiologists. FPSnet produced the lowest accuracy on this dataset, and was significantly worse than MRRN-DS in both Dataset 1 and Dataset 3. Taken together, MRRN and its variant using deep supervision (MRRN-DS) indicate potential for generating auto-segmentation of DIL as a starting contouring solution for radiologists and radiation oncologists. Further improvements may be possible with the ensembling of well performing networks, or through the incorporation of prostate segmentations as a postprocessing step (particularly for Dataset2 where there were out of prostate segmentations). The improvement in performance of FPSnet-SL vs FPSnet also motivates further research into enhancing the generalizability of PCa segmentations with label softening.

This study has the following limitations. First, a side-by-side comparison of prior methods for PCa segmentation was not possible because of large differences in the datasets used in the analysis. Second, we only used ADC instead of combining both T2-weighted and ADC images in order to avoid any errors in spatial alignment of the two images used as input to the models. Combination of the two MR sequences is planned future work. Third, the inter-rater study was performed using a small number of scans as the goal of the analysis was to assess feasibility to reduce inter-rater variability. Finally, we also did not study the potential of these methods to reduce contouring effort as this is being addressed in future research. Nevertheless, to our best knowledge, the first study represents a systematic evaluation of generalization ability of deep networks to segment PCa with variabilities stemming from cancer aggressiveness, size, location, as well as MR imaging differences.

**Table 3. Comparison of MRRN-DS against representative published PCa segmentation methods, including achieved accuracies, training and testing data configurations.**

| Method | Data description | Goal | Key results | Notes |
|---|---|---|---|---|
| MRRN-DS | 355 patients with GE and non-GE scanners and with different image acquisitions. **Training: Internal** 151 patients acquired with 1.5 Tesla GE scanner with endorectal coil **Testing**: Internal 26 patients acquired with 1.5 Tesla GE scanner with endorectal coil External 204 patients from 3.0 Tesla Siemens (Magnetom Trio and Skyra) Internal 10 patients from 3.0 Tesla Phillips MRI. | **Primary**: Segmentation of dominant index lesions (GS $\geq$3+4) **Secondary**: Segmentation of all intra-prostatic lesions including GS < 6) and varying sizes. **Tertiary**: Accuracy comparison of lesions in peripheral vs. transition zone. | DSC of 0.60 on internal Dataset 1 and 0.54 on external Dataset 2 testing sets DSC of 0.45 with respect to radiologist vs. inter-rater DSC of 0.41 | Median lesion size of GS $\geq$3+4 lesions: Internal 1.1 cc; External 1.1 cc Analyzed scans: MRI |
| Duran (2022), Multi-task attention Unet [37] | 219 MRI from three different scanners **Training:** Five-fold cross validation on 219 datasets **Testing:** ProstateX dataset | **Primary**: Prostate lesion segmentation reported using Cohen's Kappa **Secondary:** Prostate gland segmentation reported using DSC | Cohen's weighted Kappa coefficient of 0.418 on five-fold cross-validation Cohen's Kappa coefficient of 0.12 on ProstateX | Cohen's Kappa, a chance corrected measure of agreement and can be problematic for unbalanced classes. [67] |

| Saha (2021), 3D CNN [17] | 2436 patients **Training**: 1584 training, 366 validation MRI from internal 3.0 Tesla Siemens (Magnetom Trio/Skyra, and Prisma) **Testing**: Internal 486 MRI from internal 3.0 Tesla Siemens External 296 MRI from 3.0 Tesla Siemens Skyra | **Primary**: Diagnosis of biopsy confirmed PCa with PIRADS > 3 or GS > 3 + 3 | Spatial congruency analysis evaluating voxel-wise detection showed Mean DSC of 0.49 on internal and mean DSC of 0.58 on external testing | Median lesion size: Internal 1.1 cc; External 1.7 cc. Analyzed scans: T2 weighted, DWI, and ADC MRI. |
|---|---|---|---|---|
| Chen (2020), Multiple Branch Unet [18] | 136 MRI from single institution with PIRADS ≥ 4. **Testing**: Subset of 28 lesions from 162 lesions in the 136 MRI. | **Primary**: Segmentation of intra-prostatic lesions | Case specific DSC of 0.63. | Analyzed scans: T2-weighted, DWI, ADC MRI. Relatively large lesions used (Figure 3). |
| Dai (2020), Mask RCNN [5] | 120 Patients from private (3 Tesla Philips) and public ProstateX (3 Tesla Siemens). **Training**: (i) Single domain public (45) vs. (ii) Mixed domain public (45 + 21) private datasets **Testing**: (i) 23 public, 42 private MR (ii) 23 public and 21 private MRI | **Primary**: Segmentation of prostate gland and intra-prostatic lesions | (i) **Single domain**: DSC of 0.59 on public and DSC of 0.38 on private sets (ii) **Mixed domain**: DSC of 0.56 on public and 0.46 on private. | Analyzed scans: T2w and ADC MRI |
| Eidex (2021), Cascaded Scoring CNN [36] | Single institution 77 T1-weighted MRI from Siemens Area MRI **Training**: Five-fold cross-validation on 44 patient MRI **Testing**: 33 MRI | **Primary**: Segment whole prostate and DIL | Mean DSC of 0.83 for lesions with DSC > 0.4 | Removes small tumors by selecting tumors with larger DSC. |
| Jung (2019), Decoder Mixup [19] | Single institution and internal 350 T2-weighted MRI **Training**: 300 MRI **Testing**: Randomly selected 50 MRI | **Primary**: Segment histopathologically confirmed PCa | Mean DSC of 0.48 on testing set | Scanner details not provided |
| Kohl (2017), Adversarial Unet [20] | Single institution 152 patients acquired using 3.0 Tesla Siemens Training and testing: 4-fold cross validation | **Primary**: Segment tumor and prostate gland zones | Cross-validation lesion DSC of 0.41 | Analyzed scans: T2w, DWI, ADC Only cancers with GS ≥ 7 used |
| Gunashekar (2022) Unet, GradCAM [62] | Single institution 137 patient MRI acquired from 1.5 Tesla and 3Tesla Siemens **Training**: 122 patients without whole mount pathology confirmation **Testing**: 15 patients with whole mount pathology confirmation | **Primary**: Segment PCa and prostate glands | Testing accuracy DSC for PCa of 0.31 Inter-rater DSC of two radiation oncologist of 0.32 | Analyzed scans: T2w MRI, ADC and synthetic high b-value (b=1400 s/mm2) |

**CONCLUSION**

MRRN-DS was generalizable to different MR testing datasets acquired using different scanners, with higher DSC than Unet. The MRRN-DS also achieved slightly higher agreement with an experienced radiologist than two radiologists. Finally, MRRN-DS was most accurate for more aggressive lesions, which are generally candidates for radiative dose ablation.

**ACKNOWLEDGMENTS**

Work is supported in part by NIH grant P30 CA008748.The authors thank the MSKCC imaging and radiation sciences (IMRAS) seed grant.